\documentclass[10pt,conference]{IEEEtran}
\usepackage{amssymb}
\usepackage{amsmath}
\usepackage{graphicx}
\usepackage{url}
\usepackage{float}
\usepackage{cite} 
\usepackage{tikz}
\usepackage{tikzscale}
\usepackage{circuitikz}
\newtheorem{remark}{Remark}
\IEEEoverridecommandlockouts
\begin{document}
\title{On the Probability of Network States with Gaussian Connectivity Functions}

\author{

\IEEEauthorblockN{Amy S. Inwood\IEEEauthorrefmark{1}, Peter J. Smith\IEEEauthorrefmark{2}, Pawel Dmochowski\IEEEauthorrefmark{3}, Carl P. Dettmann\IEEEauthorrefmark{4}, \\ Justin P. Coon\IEEEauthorrefmark{5},~and Michail Matthaiou\IEEEauthorrefmark{1}}

\IEEEauthorblockA{\IEEEauthorrefmark{1}Centre for Wireless Innovation (CWI), Queen’s University Belfast, Belfast BT3 9DT, U.K.}

\IEEEauthorblockA{\IEEEauthorrefmark{2}School of Mathematics and Statistics, Victoria University of Wellington, Wellington, New Zealand}

\IEEEauthorblockA{\IEEEauthorrefmark{3}School of Engineering and Computer Science, Victoria University of Wellington, Wellington, New Zealand}

\IEEEauthorblockA{\IEEEauthorrefmark{4}School of Mathematics, University of Bristol, Bristol BS8 1UG, U.K.}

\IEEEauthorblockA{\IEEEauthorrefmark{5}Department of Engineering Science, University of Oxford, Oxford OX1 2JD, U.K.}

\IEEEauthorblockA{Emails: \{a.inwood, m.matthaiou\}@qub.ac.uk, \{peter.smith, pawel.dmochowski\}@vuw.ac.nz, \\ carl.dettmann@bris.ac.uk, justin.coon@eng.ox.ac.uk}

\thanks{This work was supported by the U.K. Engineering and Physical Sciences Research Council (EPSRC) grant (EP/X04047X/2) for TITAN Telecoms Hub. The work of P. J. Smith was supported by the Marsden Fund Council from New Zealand Government funding, managed by Royal Society Te Apārangi. The work of M. Matthaiou was supported by the European Research Council (ERC) under the European Union’s Horizon 2020 research and innovation programme (grant agreement No. 101001331).}}

\maketitle
\thispagestyle{empty}
\pagestyle{empty}

\begin{abstract}
In this paper, we consider the connectivity of a random network of $N$ mobile devices in three dimensions (3D), where the location of each device or node has a Gaussian distribution in each dimension. For each pair of nodes, the probability of connectivity is related to the nodes' separation by a Gaussian connectivity function. The fundamental analytical tool for studying such systems is the probability of a given network state, derived and expressed in terms of its graph Laplacian. Leveraging this result, we obtain results for the connectivity of small networks, the probability of a complete network (where all nodes are connected to all other nodes), and the probability of an isolated node, which gives an approximation to the connectivity probability of larger networks. The general results are then simplified for special cases and limiting scenarios. 
\end{abstract}


\section{Introduction}\label{intro}

The connectivity of ad hoc networks has been studied for many different scenarios. These include varying numbers of dimensions \cite{coon_full_2012}, different fading environments \cite{orestis_network_2014}, different connectivity probabilities between nodes \cite{dettmann_random_2016} and different spatial assumptions concerning node positioning \cite{dettmann_random_2016, smith_control_2023}. Simulations allow arbitrarily complex scenarios to be assessed numerically, but analytical progress is necessarily restricted to certain tractable network problems. For example, many studies assume a Poisson process for node locations \cite{coon_full_2012,coon_impact_2012},  focus on two dimension (2D) problems \cite{mao_on_2011}, assume fixed connection probabilities between nodes \cite{krishnamachari_phase_2001}, or only evaluate certain network states (e.g. connected networks) \cite{coon_full_2012,coon_impact_2012,mao_on_2011}. There are very few closed-form results available for the probability of arbitrarily connected networks when the nodes are distributed in a non-uniform manner. This forms the motivation for this work.

We adopt the empirically-validated spatial model proposed in \cite{smith_3d_2020,smith_control_2023}, where nodes are randomly distributed in three dimension (3D) space according to a Gaussian distribution. This allows node clusters to be centered at different locations with differing spreads, capturing more realistic and heterogeneous topologies than a uniform distribution, while also allowing outlier nodes that are widely separated from the rest. For networks of arbitrary size, we seek to compute the probability that the network is in any particular state, i.e. the probability that an $N$-node network is in any one of the $2^{N(N-1)/2}$ possible states, ranging from no nodes connected to all $N$ nodes connected. This enables the tractable computation of key network probabilities, such as the probability of the network being connected, or complete, where all nodes are connected to all other nodes. In most cases, the desired generality leads to a non-tractable problem. However, for the important case of Gaussian connectivity functions \cite{kartun-giles_counting_2018}, closed-form expressions can be obtained. Under this assumption, the resulting probabilities are conveniently defined in terms of the Laplacian of the network \cite{chung_spectral_1997}.

The massive number of connected networks for large $N$ implies that it is prohibitive to evaluate the probability of connectivity via an enumeration of all connected networks. Hence, when $N$ is large, we adopt the \textit{isolated node approximation} \cite{walters_random_2011} where the probability of not being connected is approximated by the probability that one or more nodes are isolated. While this approximation has been proven for networks in bounded regions \cite{penrose_random_2003,penrose_connectivity_2016}, our results demonstrate its accuracy in the non-bounded networks considered. Finally, we consider the important special case of homogeneous networks and the limiting cases of weakly and strongly connected networks.

\textit{Notation}: The probability of event $A$ is denoted $P(A)$; statistical expectation is $\mathbb{E}[\cdot]$; $\mathcal{N}(\mu,\sigma^2)$ is the Gaussian distribution with mean $\mu$ and variance $\sigma^2$; upper and lower boldface letters represent matrices and vectors, respectively; $(\mathbf{X})_{ij}$ denotes the ($i,j$)-th element of $\mathbf{X}$; $\det(\cdot)$ is the matrix determinant; $(\cdot)^T$ and $(\cdot)^{-1}$ are the transpose and inverse operators, and $\mathbf{I}_N$ is the identity matrix of order $N$.

\section{Network Connectivity} \label{conn}

\begin{figure}[t]
    \centering
    \resizebox{0.375\textwidth}{!}{\includegraphics{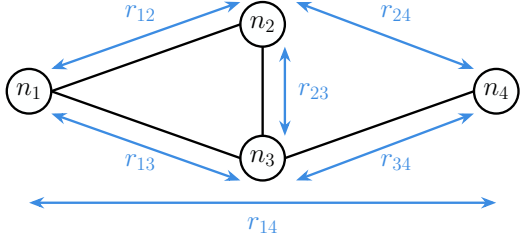}}
    \caption{Example network $G_1$. Solid black lines indicate connected links.}
    \label{fig:examplenetwork}
\end{figure}

Consider a network consisting of $N$ randomly located nodes in 3D space. For any pair of nodes, $n_i$ and $n_j$, it is assumed that the probability of connection between them follows the Gaussian connectivity function, such that
\begin{equation}\label{eq:Pij}
    P(n_i\text{ connected to }n_j) = H_{ij} = \exp\left(-r_{ij}^2/r_0^2\right),
\end{equation}
where $r_{ij}$ is the nodes' separation, and $r_0$ is a scaling constant. This formulation is widely used in fundamental studies \cite{dettmann_random_2016,kartun-giles_counting_2018}, with the parameter $r_0$ controlling the level of connectivity. As $r_0 \to 0$, connectivity reduces and as $r_0 \to \infty$, connectivity increases. Hence, $r_0$ can be viewed as a proxy for the combined effect of physical factors, such as transmit power, path loss, the medium of communication, etc. The probability of a particular network can be expressed solely in terms of products of the probabilities, $H_{ij}$, as shown in Fig. \ref{fig:examplenetwork} for the example network, $G_1$.
 As we are interested in the average probability of a particular network state over all spatial node configurations, the probability of $G_1$ is given by 
\begin{equation}    \label{eq:PN1a}
    P(G_1) = \mathbb{E}\left[H_{12}H_{13}H_{23}H_{34}(1-H_{14})(1-H_{24}) \right],
\end{equation}
where the connectivity probabilities $H_{12}, H_{13}$, etc. are conditioned on the nodes' locations, while the expectation averages over all possible locations. Expanding \eqref{eq:PN1a} gives
\begin{align} \label{eq:PN1b}
    \!P(&G_1) {=} \mathbb{E}\left[H_{12}H_{13}H_{23}H_{34} \right]{+}\mathbb{E}\left[H_{12}H_{13}H_{23}H_{34} H_{14}H_{24}\right]  \notag \\ {-}&\mathbb{E}\left[H_{12}H_{13}H_{23}H_{34}H_{24} \right]{-}\mathbb{E}\left[H_{12}H_{13}H_{23}H_{34} H_{14}\right] \!.\!
 \end{align}
Hence, the probability of the network can be obtained by evaluating terms of the form 
\begin{equation}\label{prod}
\mathbb{E}\left[ \prod_{r=1}^m \!H_{i_r j_r} \right],
\end{equation}
where
$\mathcal{S}\!=\!\left\{(i_1,j_1),\ldots,(i_m,j_m) \right\}$ is a set containing the $m$ distinct pairs of nodes in the product. For compact  notation, the dependence of the set, $\mathcal{S}$, on the $m$ nodes is omitted and inferred from the context. We also denote by $L$ the total number of nodes involved in the $m$ pairs.

The locations of the nodes are assumed to be mutually independent, while the $x,y,$ and $z$ coordinates for a given node are also independent. Let the location of $n_i$ be $(X_i, Y_i, Z_i)$, where $X_i\sim \mathcal{N}(\mu_{x_i},\sigma^2_{x_i}), Y_i\sim \mathcal{N}(\mu_{y_i},\sigma^2_{y_i}),$ and $Z_i\sim \mathcal{N}(\mu_{z_i},\sigma^2_{z_i})$. Using the independence properties and expanding \eqref{prod} gives
\begin{align}
\mathbb{E}\!\bigg[\prod_{r=1}^{m}\! H_{i_r,j_r}\! \bigg]\! &{=}\mathbb{E}\!\bigg[ \prod_{r=1}^{m}\! 
                    \exp\!\left(\!\!\tfrac{ -(X_{i_r}\!-\!X_{j_r})^2  \!- (Y_{i_r}\!-\!Y_{j_r})^2\! -(Z_{i_r}\!-\!Z_{j_r})^2\!}{r_0^2}  \right)  \!\!\!\bigg]\!,              \notag \\
                    &{=} \mathbb{E}\!\bigg[\prod_{r=1}^{m}H_{i_rj_r}^x \bigg]    \mathbb{E}\!\bigg[\prod_{r=1}^{m}H_{i_rj_r}^y \bigg]\mathbb{E}\!\bigg[\prod_{r=1}^{m}H_{i_rj_r}^z \bigg]\!,           \notag \\
                    &\!\triangleq T_x(\mathcal{S})T_y(\mathcal{S})T_z(\mathcal{S}),  \label{eq:EprodHij}
\end{align}
where $H_{ij}^x=\exp(-(X_i-X_j)^2/r_0^2)$, and $H_{ij}^y$ and $H_{ij}^z$ are defined similarly.

\subsection{Probability of a Set of Connected Nodes}\label{sec_partic}

From \eqref{eq:PN1b} and \eqref{eq:EprodHij}, it is sufficient to determine $T_x(\mathcal{S})$ in order to compute any network probability, since $T_y(\mathcal{S})$ and $T_z(\mathcal{S})$ can be obtained analogously. We therefore derive $T_x(\mathcal{S})$ below:
\begin{align}
    T_x(\mathcal{S})=&\,\mathbb{E}\bigg[\prod_{r=1}^{m} \exp\! \left(\!\tfrac{ -(X_{i_r}-X_{j_r})^2 }{r_0^2}\right)\!\! \bigg], \notag \\
    =&\int_{-\infty}^{\infty} \cdots \int_{-\infty}^{\infty} 
    \prod_{r=1}^{m} \exp \left( \tfrac{ -(x_{i_r}-x_{j_r})^2 }{r_0^2} \right) \notag \\ &\qquad\times\prod_{t=1}^{L} \tfrac{1}
     {\sqrt{2\pi \sigma^2_{x_{k_t}}}}\exp \bigg( \tfrac{ -\left(x_{k_t}-\mu_{x_{k_t}}\right)^2 }{2\sigma^2_{x_{k_t}}}\bigg) d \mathbf{x}, \label{eq:Txs}
\end{align}
where $\mathbf{x}=[x_{k_1},x_{k_2},\ldots,x_{k_L}]$, while $k_1$ to $k_L$ index the $L$ different nodes involved in set $\mathcal{S}$. Converting \eqref{eq:Txs} to a single exponential term gives
\begin{equation}
    T_x(\mathcal{S}){=}\frac{\int_{-\infty}^{\infty}\! \!\cdots\!\! \int_{-\infty}^{\infty}\!\exp\!{\Big(\!\!{\scriptstyle-\!\!\!\sum\limits_{r=1}^{m} \!\!\frac{(x_{i_r}\!-x_{j_r})^2 }{r_0^2}-\!\!\sum\limits_{t=1}^{L}\!\! \frac{ (x_{k_t}\!-\mu_{x_{k_t}}\!)^2 }{2\sigma^2_{x_{k_t}}}}\!\Big)} d\mathbf{x}}{(2\pi)^{L/2} \prod_{t=1}^{L}\sigma_{x_{k_t}}}. \!\label{eq:Txs2}
\end{equation}
Completing the square in \eqref{eq:Txs2} allows $T_x(\mathcal{S})$ to be written as
\begin{align}
    T_x(\mathcal{S})=&\frac{
             \int_{-\infty}^{\infty} \!\!\cdots\!\! \int_{-\infty}^{\infty}\!
             \mathrm{e}^{
             -\frac{1}{2}
             \left( \mathbf{x}-\mathbf{a}_x(\mathcal{S})  \right)^T {\boldsymbol{\Sigma}}_x^{-1}(\mathcal{S})  \left( \mathbf{x}-\mathbf{a}_x(\mathcal{S})   \right)}\mathrm{e}^{C_x(\mathcal{S})}d\mathbf{x}}{(2\pi)^{L/2} \prod_{t=1}^{L}\sigma_{x_{k_t}}}, \notag \\
         =&e^{C_x(\mathcal{S})} \det({\boldsymbol{\Sigma}}_x(\mathcal{S}))^{1/2} \prod_{t=1}^{L}\frac{1}{\sigma_{x_{k_t}}},
\end{align}
using properties of the multivariate normal density. The mechanics of completing the square to derive $\mathbf{a}_x(\mathcal{S})$, ${\boldsymbol{\Sigma}}_x(\mathcal{S})$ and $C_x(\mathcal{S})$ are lengthy but straightforward. Hence, only the results are given here. Defining $\mathbf{M}_x(\mathcal{S})=\boldsymbol{\Sigma}_x^{-1}(\mathcal{S})$ we have
\begin{align}
    T_x(\mathcal{S})=&\frac{ e^{C_x(\mathcal{S})} }
         { \det(\mathbf{M}_x(\mathcal{S}))^{1/2} \prod_{t=1}^{L}\sigma_{x_{k_t}}  },
         \label{eq:Txs4}
\end{align}
where
\begin{equation}
    C_x(\mathcal{S})=\frac{1}{2}
         \bigg(\mathbf{a}^T_x(\mathcal{S}) \mathbf{M}_x(\mathcal{S}) \mathbf{a}_x(\mathcal{S}) - \sum_{t=1}^{L}\frac{\mu^2_{x_{k_t}}}{\sigma^2_{x_{k_t}}}\bigg), 
\end{equation}         
\begin{equation}
    \mathbf{a}^T_x(\mathcal{S})=\bigg[ \frac{\mu_{x_{k_1}}}{\sigma^2_{x_{k_1}}},\ldots, \frac{\mu_{x_{k_L}}}{\sigma^2_{x_{k_L}}}\bigg] \mathbf{M}_x^{-1}(\mathcal{S}) ,
\end{equation}
and 
\begin{align}
    \!\left(\mathbf{M}_x(\mathcal{S})\!\right)_{rr}&{=}1/\sigma^2_{x_{k_r}}\!+2\nu_{k_r}/r_0^2,  \label{eq:MxSrr} \\
    \!\left(\mathbf{M}_x(\mathcal{S})\!\right)_{rq}&{=}\!
    \begin{cases}
        0 & \text{\!\!\!\!if $n_{k_r}\!$ and $n_{k_q}\!$ are not connected,} \\
        \!-2/r_0^2 & \text{\!\!\!\!if $n_{k_r}\!$ and $n_{k_q}\!$ are connected,}
    \end{cases}
    \label{eq:MxSrq}
\end{align}
where $\nu_{k_r}$ is the number of connections from $n_{k_r}$ in the set $\mathcal{S}$. Note that $\mathbf{M}_x(\mathcal{S})$ can be expressed in terms of the Laplacian of $\mathcal{S}$. From \eqref{eq:MxSrr} and \eqref{eq:MxSrq}, $\mathbf{M}_x(\mathcal{S})$ can be written as
\begin{align}
    \mathbf{M}_x(\mathcal{S})&= \mathrm{diag}\!\left[\! \tfrac{1}{\sigma^2_{x_{k_1}}\!},\ldots,\!\tfrac{1}{\sigma^2_{x_{k_L}}\!}\!  \right]
    +\tfrac{2}{r_0^2}(\mathbf{D}(\mathcal{S})-\mathbf{A}(\mathcal{S})), \notag \\
    &= \mathrm{diag}\!\left[\! \tfrac{1}{\sigma^2_{x_{k_1}}\!},\ldots,\!\tfrac{1}{\sigma^2_{x_{k_L}}\!}\!  \right]
    +\tfrac{2}{r_0^2}\mathbf{L}(\mathcal{S}), \label{eq:M}
\end{align}
where $\mathbf{D}(\mathcal{S})=\mathrm{diag}[\nu_{k_1},\ldots,\nu_{k_L}]$ is the degree matrix, $\mathbf{A}(\mathcal{S})$ is the adjacency matrix (containing 1 in positions of connected pairs) and $\mathbf{L}(\mathcal{S})$ is the Laplacian of $\mathcal{S}$ \cite{chung_spectral_1997}.

\section{Special Cases} \label{cases}
\subsection{Complete Network} \label{Sec:full}

In a complete network, all $N$ nodes are connected to all other nodes and the connectivity probability is defined by 
\begin{equation}
    P_{f}=\mathbb{E} \Bigg[\prod_{i=1}^{N-1} \prod_{j=i+1}^{N} H_{ij} \Bigg].
    \label{eq:Pfull}
\end{equation}
Note that \eqref{eq:Pfull} is a special case of \eqref{eq:EprodHij}, where $\mathcal{S}=\left\{ (1,2), (1,3), (2,3), (2,4),\dots,((N-1),N) \right\}$. This can be computed using \eqref{eq:EprodHij}, with $T_x(\mathcal{S})$ as defined in \eqref{eq:Txs4}.

\subsection{Connected Network} \label{Sec:connected} 

\begin{figure}[t]
    \centering
    \resizebox{0.45\textwidth}{!}{\includegraphics{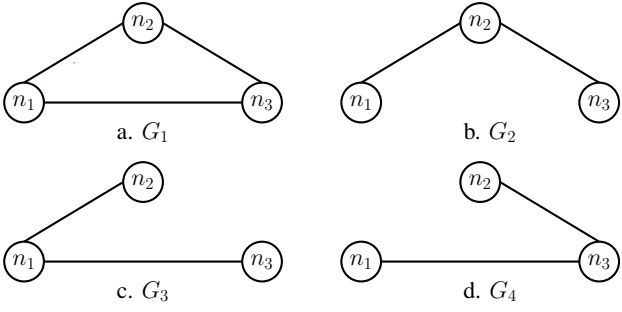}}
    \caption{The four possible connected configurations of a three-node network.}
    \label{fig:threenodenet}
\end{figure}

An exact calculation of connectivity requires the enumeration of all possible connected networks. Figure \ref{fig:threenodenet} shows the four configurations of a three-node connected network.
Thus, the connection probability of a three-node network is
\begin{align}
    P_c&=\sum_{i=1}^{4} P(G_i), \notag \\
    &=\mathbb{E} \left[H_{12}H_{13}H_{23}\right]+\mathbb{E} \left[H_{12}(1-H_{13})H_{23}\right]\notag \\
     &\qquad+\!\mathbb{E} \left[H_{12}H_{13}(1{-}H_{23})\right]\!{+}\mathbb{E} \left[(1{-}H_{12})H_{13}H_{23}\right]\!, \notag \\
   & =\mathbb{E} \left[H_{12}H_{23}\right]+\mathbb{E}\left[H_{12}H_{13}\right] \notag \\ &\qquad\qquad\qquad\,+\mathbb{E}\left[H_{13}H_{23}\right]-2\mathbb{E}\left[H_{12}H_{13}H_{23}\right].\label{eq:Pc}
\end{align}
In general, for an arbitrary $N$-node network, $P_c=\sum_{i=1}^{N_c} P(G_i) $, where $N_c$ is the number of possible connected networks and each $P(G_i)$ can be computed using the methodology in Section \ref{sec_partic}. For $N=3$, there are $N_c=4$ possibilities, for $N=4$ there are $N_c=38$ possibilities, and for $N=5$ there are $N_c=728$ possibilities \cite{harary_graphical_1973}. Beyond $N=5$, enumeration of the connected networks is cumbersome and we resort to approximate results based on isolated node probabilities, as detailed in Section \ref{sec_NI}.

\subsection{Probability of an Isolated Node}\label{sec_NI}
An approximation to $P_c$ can be constructed by identifying the presence of an isolated node as the dominant mechanism for a lack of connectivity \cite{walters_random_2011}. By neglecting the dependence between isolation events, as this is known to be small \cite{penrose_random_2003}, the connection probability of an $N$ node network can be written
\begin{align}
P_c &\approx \prod_{k=1}^N (1-P_{I,k}), \label{eq:Pcisolated} \\
&= \prod_{k=1}^N\Bigg(1-\sum_{r=0}^{N-1}(-1)^r 
\sum_{\mathcal{S}_{rk}} 
\mathbb{E}\!\Bigg[ \prod_{j=1}^{r} H_{k i_j} \Bigg]\Bigg),
\label{eq:Pcisolated3}
\end{align}
where $P_{I,k} $ is the probability that $n_k$ is isolated and $\mathcal{S}_{rk}=\left\{ i_1,\ldots,i_r:1 \le i_1 < \cdots < i_r \le N,\; i_j \neq k \right\}.$ 
The value of  $P_{I,k} $ used in \eqref{eq:Pcisolated3} is established using the inclusion-exclusion principle. This approximation assumes that the events ``node $k$ is not isolated'' are mutually independent. This is not strictly true, particularly in small networks where the isolation of one node strongly affects the connectivity probability of its neighbors. However, the dependence weakens rapidly as $N$ increases. To illustrate this, consider the number of labeled graphs in which a given node is connected. For $N=3$, there are 6 such configurations, for $N=4$, there are 56, and for $N=5$, there are 960 \cite{harary_graphical_1973}. As $N$ increases, the occurrence of one isolated node has an increasingly negligible influence on the isolation probability of others due to the proliferation of connected states. Consequently, joint isolation events become  weakly dependent, explaining the remarkable accuracy of the product approximation, as demonstrated in Section \ref{results}.


\section{Homogeneous Networks} \label{homogeneous}
We now consider the homogeneous case, where the position of each node is zero-mean, independent and identically distributed, i.e. the co-ordinates corresponding to $n_i$ are $X_i\sim\mathcal{N}(\mu,\sigma^2)$, $Y_i\sim\mathcal{N}(\mu,\sigma^2)$ and $Z_i\sim\mathcal{N}(\mu,\sigma^2)$. In this case, \eqref{eq:Txs4} collapses to
\begin{equation}
T_x(\mathcal{S})=T_y(\mathcal{S})=T_z(\mathcal{S})=\sigma^{-L}\det(\mathbf{M}_x(\mathcal{S}))^{-1/2},
\end{equation}
where
\begin{equation}
    \mathbf{M}_x(\mathcal{S})=\tfrac{1}{\sigma^2}\mathbf{I}_L+\tfrac{2}{r_0^2}\mathbf{L}(\mathcal{S}), \label{eq:MxShomo}
\end{equation}
which is a simplified version of \eqref{eq:M}. Hence, the general term is given by
\begin{equation}
    \mathbb{E}\left[\prod_{r=1}^mH_{i_r,j_r}\right]=\sigma^{-3L}\det(\mathbf{M}_x(\mathcal{S}))^{-3/2}. \label{eq:prodhijhomo}
\end{equation}
In the limiting case $r_0\!\to\! 0$, $\det(\mathbf{M}_x(\mathcal{S})\!)\sim(2/{r_0^2})^L\!\det(\mathbf{L}(\mathcal{S})\!)$, so
\begin{equation}
    \mathbb{E}\left[\prod_{r=1}^mH_{i_r,j_r}\right]\!\sim\!\left(\frac{r_0}{\sigma\sqrt{2}}\right)^{\!\!3L}\!\!\det(\mathbf{L}(\mathcal{S}))^{-3/2},\,\,\, \text{as } r_0\to 0. \label{eq:genhomolim0}
\end{equation}
\begin{remark}
\label{rem:EHij_r0to0}
    For small $r_0$, the expectation is proportional to $r_0^{\,\,3L}$, where $L$ is the number of nodes involved in the product.
\end{remark}

When $r_0\to\infty$, we use the result 
\begin{equation}
    \det(\mathbf{I}_L+\mathbf{E})\sim 1+\sum_{r=1}^L(\mathbf{E})_{rr}\quad \text{as } (\mathbf{E})_{ij}\to0\,\, \forall i,j, \label{eq:matlemma}
\end{equation}
for an $L\times L$ matrix, $\mathbf{E}$, with small magnitude elements. Hence,
\begin{equation}
    \det(\mathbf{M}_x(\mathcal{S}))\sim\left(\frac{1}{\sigma^2}\right)^{\!\!L}\bigg(1+\sum_{r=1}^{L}\frac{2\sigma^2}{r_0^2}(\mathbf{L}(\mathcal{S}))_{rr}\bigg).
\end{equation}
The diagonal of $\mathbf{L}(\mathcal{S})$ is the degree matrix, 
\begin{equation}
    \sum_{r=1}^L(\mathbf{L}(\mathcal{S}))_{rr}=\sum_{r=1}^L\nu_{k_r}=\nu_s,
\end{equation}
where $\nu_s$ is the sum of the number of connections at each node. Hence, we obtain
\begin{equation}
    \mathbb{E}\left[\prod_{r=1}^mH_{i_r,j_r}\right]\sim\left(1+\frac{2\nu_s\sigma^2}{r_0^2}\right)^{-3/2} \quad \text{as } r_0\to\infty. \label{eq:genhomoliminfty}
\end{equation}
\begin{remark}
    \label{rem:EHij_r0toinf}
    Expanding \eqref{eq:genhomoliminfty} gives asymptotic equivalence with $1-{3\nu_s\sigma^2}/{r_0^2}$. Unlike Remark \ref{rem:EHij_r0to0}, the expectation as $r_0$ becomes large is a function of $r_0^{\,\,-2}$, and the power is independent of the number of nodes.
\end{remark}

\subsection{Complete Network}
For a complete network \cite{chung_spectral_1997},
\begin{equation}
    \mathbf{L}(\mathcal{S})={\scalebox{0.7}{$\begin{bmatrix}
        N-1 & -1 & \cdots & -1 \\
        -1 & N-1 & & \vdots \\
        \vdots & & \ddots & -1 \\
        -1 & \cdots & -1 & N-1
    \end{bmatrix}$}},
\end{equation}
and using \eqref{eq:MxShomo} gives
\begin{equation}
    \mathbf{M}_x(\mathcal{S}) = \left(\frac{1}{\sigma^2}+\frac{2N}{r_0^2}\right)\mathbf{I}_N - \frac{2}{r_0^2}.
\end{equation}
Using the matrix determinant lemma \cite{ding_eigenvalues_2007},
\begin{equation}
    \det(\mathbf{M}_x(\mathcal{S})) = \left(\frac{1}{\sigma^2}+\frac{2N}{r_0^2}\right)^N\left(1-\frac{2N/r_0^2}{\frac{1}{\sigma^2}+\frac{2N}{r_0^2}}\right). \label{eq:detMhomo}
\end{equation}
Substituting \eqref{eq:detMhomo} into \eqref{eq:prodhijhomo} and simplifying gives the probability of a complete homogeneous network as
\begin{equation}
    P_f = \left(1+\frac{2N\sigma^2}{r_0^2}\right)^{-3(N-1)/2}. \label{eq:Pfhomo}
\end{equation}
Considering the limiting cases, it is straightforward from \eqref{eq:Pfhomo} to show that 
\begin{align}
    P_f&\sim(r_0/(\sigma\sqrt{2N}))^{3(N-1)}  &&\text{as } r_0\to0.\\
    P_f&\sim1-3N(N-1)\tfrac{\sigma^2}{r_0^2} \quad &&\text{as } r_0\to\infty.
\end{align}
These results can  be found via \eqref{eq:genhomolim0} and \eqref{eq:genhomoliminfty}.
\begin{remark}
    \label{rem:Pf_lims}
    A direct implication of Remarks \ref{rem:EHij_r0to0} and \ref{rem:EHij_r0toinf} is the dependence of $P_f$ on $r_0^{\,\,3(N-1)}$ and $r_0^{-2}$ for $r_0\to0$ and $r_0~\to~\infty$, respectively.
\end{remark}

\subsection{Isolated Node}
As in the general case, the probability of a connected network can be approximated using the probability of an isolated node as in \eqref{eq:Pcisolated}. In the homogeneous case, the expression for $P_{I,k}$ simplifies significantly. When the positions of all nodes are independent and identically distributed (i.i.d.) random variables, the expectation $\sum_{\mathcal{S}_{rk}}\mathbb{E}\left[\prod_{j=1}^rH_{ki_j}\right]$ depends only on the number of nodes in $\mathcal{S}_{rk}$, rather than which nodes $\mathcal{S}_{rk}$ contains. This means that all subsets of size $r$ have the same expectation. Therefore, a binomial coefficient can be factored out, and $P_{I,k}$ becomes
\begin{align}
    P_{I,k}&\!=\!1\!+\!\sum_{r=1}^{N\!-\!1}\!(-1)^r\!\binom{N\!-\!1}{r}\mathbb{E}\!\Bigg[\prod_{j=1}^rH_{kj}\Bigg], \notag \\
    &\!=\!1\!+\!\sum_{r=1}^{N\!-\!1}(-1)^r\!\binom{N-1}{r}\sigma^{-3r}\!\det\left(\mathbf{M}_x(\mathcal{S}_{rk})\!\right)^{-3/2}\!,
\end{align}
where, dependent on subset size $r$, $\mathbf{M}_x(\mathcal{S}_{rk})=\frac{1}{\sigma^2}\mathbf{I}_{r+1}+\frac{2}{r_0^2}\mathbf{L}(\mathcal{S}_{rk})$. Here, $\mathbf{L}(\mathcal{S}_{rk})$ is the star graph Laplacian \cite{chung_spectral_1997}
\begin{equation}
    \mathbf{L}(\mathcal{S}_{rk})={\scalebox{0.7}{$\begin{bmatrix}
        r & -1 & -1 &\cdots & -1 \\
        -1 & 1 & 0 & \cdots & 0\\
        -1 & 0 & 1 & & 0 \\ 
        \vdots & \vdots & & \ddots & \\
        -1 & 0 & 0 & \cdots & 1
    \end{bmatrix}$}},
\end{equation}
which has the known eigenvalues 
\begin{equation}
    \lambda_j= \begin{cases}
        0 & j=1 \\
        1 &2\leq j \leq r \\
        r + 1 \quad& j = r+1
    \end{cases}.
\end{equation}
This leads to
\begin{align}
    \!\!\det(\mathbf{M}_x(\mathcal{S}_{rk}))&=\prod_{j=1}^{r+1}\left(\frac{1}{\sigma^2}+\frac{2}{r_0^2}\lambda_j\right), \notag \\ 
    & =\frac{1}{\sigma^2}\!\left(\frac{1}{\sigma^2}+\frac{2}{r_0^2}\right)^{\!\!r-1}\!\!\!\left(\frac{1}{\sigma^2}+\frac{2(r+1)}{r_0^2}\!\right)\!.
\end{align}
Therefore, 
\begin{multline}
    P_{I,k}=\!1\!+\!\sum_{r=0}^{N-1}(-1)^r\binom{N-1}{r}\sigma^{-3r} \\ \times\bigg(\frac{1}{\sigma^2}\!\left(\frac{1}{\sigma^2}{+}\frac{2}{r_0^2}\right)^{\!\!r-1}\!\!\!\left(\frac{1}{\sigma^2}{+}\frac{2(r+1)}{r_0^2}\!\right)\!\bigg)^{-3/2}. \label{eq:PIkhomo}
\end{multline}
Expanding \eqref{eq:PIkhomo}, some lengthy algebra identifies the leading terms in the expansion giving the limiting cases:
\begin{align}
P_{I,k}&\sim 1-\frac{(N-1)r_0^3}{8\sigma^3} \quad \text{as } r_0\to 0, \label{first_asy}\\ 
P_{I,k}&\sim\!\!\left(\!\frac{\sigma^2}{2r_0^2}\!\right)^{\!\!N-1}\sum_{j=0}^{N-1}\!\!\frac{(N\!-\!1)!(2N\!-\!2j\!-\!1)!}{j!((N\!-\!j\!-\!1)!)^2} \quad \text{as } r_0\to \infty.\label{second_asy}
\end{align}
\begin{remark}
    \label{rem:PI_lims}
    Substituting \eqref{first_asy} and \eqref{second_asy} into \eqref{eq:Pcisolated} shows that $P_c$ is proportional to $r_0^{\,\,3N}$ for small $r_0$, and that $1-P_c$ is proportional to $r_0^{\,\,-2(N-1)}$ for large $r_0$.
\end{remark}
\section{Numerical results} \label{results}
This section verifies the analytical results and explores the probability of connectivity in a range of scenarios. For all simulations, $10^5$ replicas are generated.

\subsection{Probability of a Network}    
\begin{figure}[t]
    \centering
    \includegraphics[width=\linewidth]{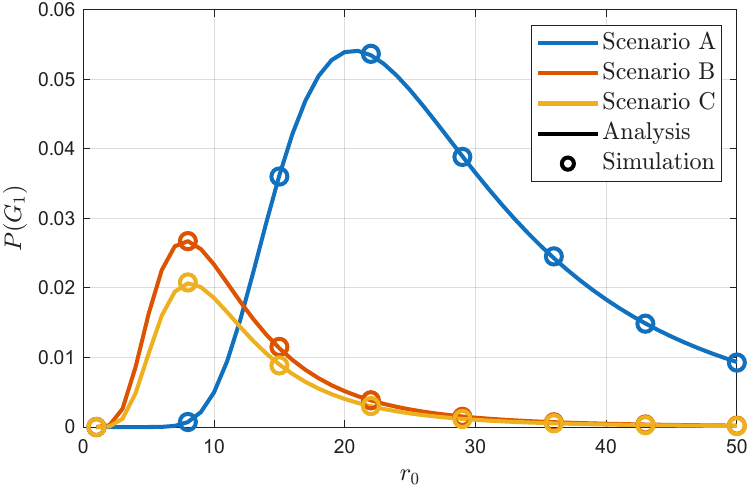}
    \caption{The probability of network $G_1$ in three different layout scenarios.}
    \label{fig:scenarios}
\end{figure}

Figure \ref{fig:scenarios} shows the probability of the four-node example network $G_1$ in Fig. \ref{fig:examplenetwork} for three different layout scenarios. Scenario A considers a heterogeneous network, where $\mu_{x_1}=\mu_{x_2}=-8$, $\mu_{x_3}=\mu_{x_4}=8$, all $\mu_{y_i}$ and $\mu_{z_i}$ values are set to 0, and all $\sigma_{x_i}$, $\sigma_{y_i}$ and $\sigma_{z_i}$ values are equal to $\sqrt{5}$. Scenario B also considers a heterogeneous network, where all $\mu_{x_i}$, $\mu_{y_i}$ and $\mu_{z_i}$ values are 0,  sigma values corresponding to nodes 1 and 2 are set to 1, and sigma values corresponding to nodes 3 and 4 are set to 3. Finally, scenario C considers a homogeneous network, where all $\mu_{x_i}$, $\mu_{y_i}$ and $\mu_{z_i}$ values are 0 and all sigma values corresponding to all nodes are equal to $\sqrt{5}$. The value of $\sqrt{5}$ is chosen so that the sum of the variances is the same in all three scenarios. The analytical results for the heterogeneous networks in scenarios A and B are computed using \eqref{eq:EprodHij}, and those for scenario C are computed using \eqref{eq:prodhijhomo}. 

Figure \ref{fig:scenarios} shows that $P(G_1)\to0$ as both $r_0\to 0$ and $r_0\to\infty$. In the low-connectivity regime ($r_0\to 0$), it is highly probable that many links are disconnected, whereas in the high-connectivity regime ($r_0\to\infty$), it is highly probable that all links are connected. Both extremes are therefore unfavorable for $P(G_1)$, which requires all links to be connected except exactly one. In Scenario A, the node positions exhibit a large gap between their mean $x$-values, which compounds the lack of connectivity already present at low $r_0$, thereby shifting the peak of $P(G_1)$ to a higher value of $r_0$. The curves corresponding to Scenarios B and C exhibit coincident peaks at the same $r_0$ value due to their identical mean node positions. However, Scenario B achieves a higher $P(G_1)$ than Scenario C, as its larger variances increase the spatial spread and thus the likelihood of exactly one disconnected link. Notably, Scenario A exhibits a significantly higher peak $P(G_1)$ than either Scenario B or C, since the inherent spatial separation between node clusters similarly promotes the single-link disconnection event characterizing $G_1$.

\subsection{Probability of a Complete Network}

\begin{figure}[t]
    \centering
    \includegraphics[width=\linewidth]{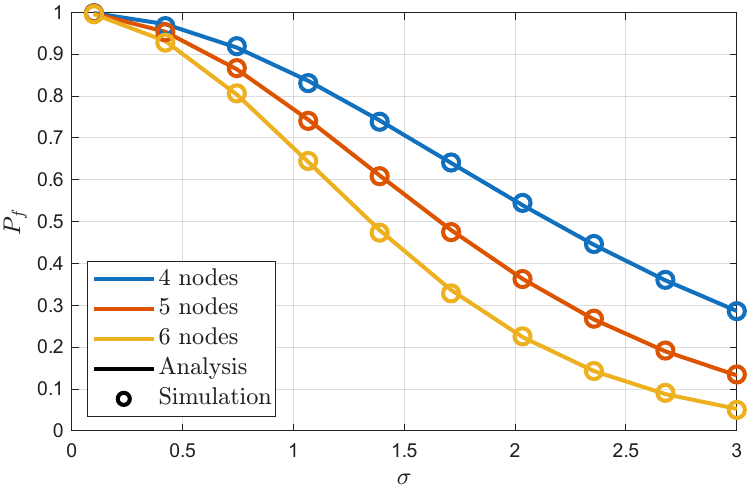}
    \caption{The probability of complete connectivity for homogeneous networks with varying $\sigma$ values.}
    \label{fig:nodessigma}
\end{figure}

Figure \ref{fig:nodessigma} investigates the probability that 4, 5 and 6 node networks are complete. Here, $r_0=15$, all $\mu$ values are set to 0, and the variance parameters are identical across all nodes. This common $\sigma$ value is varied throughout the simulation. Analytical results are calculated using \eqref{eq:Pfhomo}.

When $\sigma=0$, $P_f=1$ for all networks considered. This occurs as the mean position of all nodes in all dimensions is 0. When $\sigma=0$, all nodes are collocated at the mean and are thus guaranteed to be connected. As the value of $\sigma$ increases, the greater spatial spread increases the likelihood of a disconnected link, causing $P_f$ to decrease. As $\sigma\to\infty$, $P_f\to 0$ for all numbers of nodes as the probability of any two being connected tends to 0. Furthermore, increasing the number of nodes reduces $P_f$, as a greater number of nodes yields a higher number of inter-node links and thereby increases the probability that at least one link is disconnected.

\subsection{Probability of Network Connectivity}   

\begin{figure}[t]
    \centering
    \includegraphics[width=\linewidth]{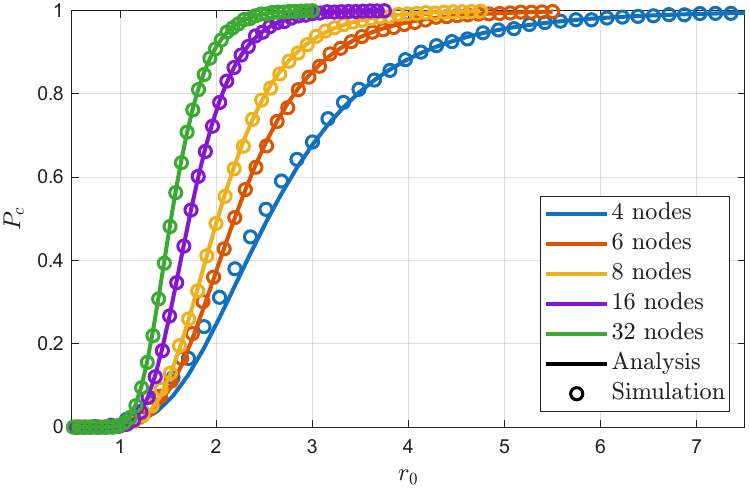}
    \caption{The probability of connectivity for homogeneous networks of varying numbers of nodes using the isolated node approximation.}
    \label{fig:isonode}
\end{figure}

Figure \ref{fig:isonode} explores the probability of network connectivity for homogeneous networks of varying numbers of nodes. The isolated node approximation in \eqref{eq:Pcisolated3} is used for analytical results, with \eqref{eq:PIkhomo} used to obtain $P_{I,k}$. This approach significantly reduces complexity and enables the study of much larger networks.

It is observed that the isolated node approximation in \eqref{eq:Pcisolated3} is remarkably accurate across the full range of $P_c$ values. The approximation is most effective in high-connectivity regimes, as in the upper tail, any lack of connectivity is more likely attributable to a single isolated node than to the formation of multiple separate connected subnetworks. Thus, this region is where analysis and simulations match most closely. The approximation also improves for larger networks, as it relies on the assumption that node isolation events are mutually independent. This assumption becomes increasingly justified as the network grows, since the disconnection of a single node has a diminishing influence on the connectivity of its neighbors. This is particularly useful in practice, as larger networks are precisely those for which enumerating all possible connected topologies becomes computationally prohibitive. Figure \ref{fig:isonode} also shows that $P_c$ increases with both $r_0$ and the number of nodes, since a greater number of nodes yields more possible connected configurations, raising the overall probability of connectivity.

\begin{figure}[t]
    \centering
    \includegraphics[width=\linewidth]{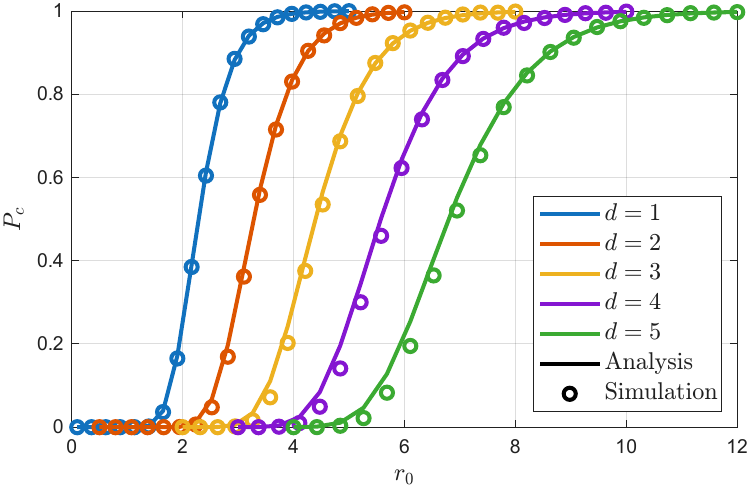}
    \caption{The probability of connectivity for 16 node heterogeneous networks using the isolated node approximation. The mean values are arranged in a grid at intervals of $d$ in both the $x$ and $y$ planes.}
    \label{fig:isonode_het}
    \vspace{-1em}
\end{figure}

Figure \ref{fig:isonode_het} investigates the probability of connectivity in 16 node heterogeneous networks. Here, the mean $x$ and $y$ co-ordinates of all nodes are arranged in a grid structure at intervals of $d$ m in each direction giving a $4 \times 4$ layout. Multiple values of $d$ are considered. Again, the isolated node approximation is used, where \eqref{eq:Pcisolated3} is used to obtain $P_c$.

Figure \ref{fig:isonode_het} shows that as $H_{ij}$ decreases exponentially with $r_{ij}$ as in \eqref{eq:Pij}, a larger separation between mean node positions necessitates a higher $r_0$ to achieve the same probability of connectivity. Figure \ref{fig:isonode_het}  also demonstrates that the isolated node approximation remains extremely accurate for heterogeneous networks, with accuracy increasing as $d$ decreases. At low values of $d$, the mean node positions are more closely spaced, reducing the likelihood of disconnection and thus increasing the probability that any disconnection is attributable to a single isolated node, which is precisely the regime in which the approximation is most valid.

\section{Conclusion}
In this paper, we investigated the connectivity of random networks 
in 3D space, where node locations followed a Gaussian distribution 
and connection probabilities were governed by a Gaussian connectivity function. A general framework using the graph Laplacian was developed for computing the probability of any network state. Leveraging this, closed-form expressions were derived for the probability of complete connectivity and the probability of an isolated node, and the latter was used to obtain an accurate and tractable approximation to the network connectivity probability. Numerical results demonstrated the remarkable accuracy of the isolated node approximation. This accuracy was shown to increase with network size and connectivity level, as the assumptions underlying the approximation become increasingly justified.

\bibliographystyle{IEEEtran}
\bibliography{references}

@article{coon_full_2012,
    author = {Coon, J. P. and Dettmann, C. P. and Georgiou, O. },
    title = {Full Connectivity: Corners, Edges and Faces},
    journal = {J. Stat. Phys.},
    month={June},
    year = {2012},
    pages = {758-778},
    volume = {147},
}

@article{coon_impact_2012,
    author = {Coon, J. P. and Dettmann, C. P. and Georgiou, O.},
    title = {Impact of boundaries on fully connected random geometric networks},
    journal = {Phys. Rev. E, Stat. Nonlin. Soft Matter Phys.},
    volume ={85},
    number={1},
    month={June},
    year = {2012},
    pages={011138-1-011138-5},
}

@INPROCEEDINGS{mao_on_2011,
  author={Mao, Guoqiang and Anderson, Brian D. O.},
  booktitle={Proc. IEEE INFOCOM}, 
  title={On the asymptotic connectivity of random networks under the random connection model}, 
  year={2011},
  pages={631-639},
  month={Apr.},}

@article{dettmann_random_2016,
  title = {Random geometric graphs with general connection functions},
  author = {Dettmann, C. P. and Georgiou, O.},
  journal = {Phys. Rev. E},
  volume = {93},
  number = {3},
  pages = {032313-1-032313-14},
  year = {2016},
  month = {Mar.},
}

@INPROCEEDINGS{smith_3d_2020,
  author={Smith, Peter J. and others},
  booktitle={Proc. IEEE PIMRC}, 
  title={{3D} Mobility Models and Analysis for {UAVs}}, 
  year={2020},
  month={Aug.},}

@incollection{chung_spectral_1997,
  author = {Chung, Fan R. K.},
  title = {Spectral {G}raph {T}heory},
  publisher = {American Mathematical Society},
  year = {1997},
  booktitle = {Regional Conference Series in Mathematics},
  address = {Providence, RI},
}

@book{harary_graphical_1973,
    author = {F. Harary and E. M. Palmer},
    title = {Graphical Enumeration},
    address = {New York, NY},
    publisher = {Academic Press},
    year = {1973}
}

@article{ding_eigenvalues_2007,
    title = {Eigenvalues of rank-one updated matrices with some applications},
    journal = {Appl. Math. Lett.},
    volume = {20},
    number = {12},
    pages = {1223-1226},
    year = {2007},
    month={Dec.},
    author = {Jiu Ding and Aihui Zhou}
}

@ARTICLE{smith_control_2023,
  author={Smith, Peter J. and Singh, Ikram and Dmochowski, Pawel and Dettmann, Carl P. and Coon, Justin Porter},
  journal={IEEE Trans. Veh. Technol.}, 
  title={Control Mechanisms for Mobile Devices}, 
  year={2023},
  month={May},
  volume={72},
  number={5},
  pages={6001-6008},
}

@incollection{walters_random_2011,
    author = {M. Walters},
    booktitle = {Surveys in Combinatorics 2011},
    publisher = {Cambridge University Press},
    year = {2011},
    title = {Random Geometric Graphs},
    address = {Cambridge, U.K.}
}

@INPROCEEDINGS{krishnamachari_phase_2001,
  author={Krishnamachari, B. and Wicker, S. B. and Bejar, R.},
  booktitle={Proc. IEEE GLOBECOM}, 
  title={Phase transition phenomena in wireless ad hoc networks}, 
  year={2001},
  pages={2921-2925},
  month={Nov.},}

@incollection{penrose_random_2003,
    author = {Matthew Penrose},
    title = {Random {G}eometric {G}raphs},
    booktitle = {Oxford Studies in Probability},
    publisher = {Oxford University Press},
    address = {Oxford, U.K.},
    year = {2003}
}

@article{penrose_connectivity_2016,
    author = {Matthew D. Penrose},
    title = {Connectivity of Soft Random Geometric Graphs},
    journal = {Ann. Appl. Probab.},
    volume={26},
    number={2},
    pages={986-1028},
    month = {Apr.},
    year = {2016}
}

@ARTICLE{kartun-giles_counting_2018,
  author={Kartun-Giles, Alexander P. and Kim, Sunwoo},
  journal={IEEE Trans. Wireless Commun.}, 
  title={Counting  $k$-Hop Paths in the Random Connection Model}, 
  year={2018},
  volume={17},
  number={5},
  pages={3201-3210},
  month={May},}

@INPROCEEDINGS{orestis_network_2014,
  author={Georgiou, Orestis and Dettmann, Carl P. and Coon, Justin P.},
  booktitle={Proc. IEEE ICC}, 
  title={Network connectivity: Stochastic vs. deterministic wireless channels}, 
  year={2014},
  pages={77-82},
  month={June},}

\end{document}